\documentclass{iau} 

\usepackage{graphicx}

\title[Correlations in pulsar variability] 
{Correlated emission and spin-down variability in radio pulsars}

\author[Benjamin Shaw]   
{Benjamin Shaw,$^1$
Benjamin W. Stappers,$^1$
Paul R. Brook,$^2$
Aris Karastergiou,$^3$
Andrew G. Lyne,$^1$
\and Patrick Weltevrede$^1$}

\affiliation{$^1$Jodrell Bank Centre for Astrophysics, \\ The University of Manchester,
Manchester, M13 9PL, UK \\ email: {\tt benjamin.shaw@manchester.ac.uk} \\[\affilskip]
$^2$Dept. of Physics and Astronomy, West Virginia University, \\ 
P.O. Box 6315, Morgantown, WV 26505, USA \\ 
$^3$Astrophysics, University of Oxford, Denys Wilkinson Building, \\
Keble Road, Oxford, OX1 3RH, UK}

\pubyear{2017}
\volume{337}  
\jname{Pulsar Astrophysics -­ The Next 50 Years}
\editors{P. Weltevrede, B.B.P. Perera, L. Levin Preston \& S. Sanidas, eds.}

\begin{document}

\maketitle

\begin{abstract}
The recent revelation that there are correlated period derivative and pulse shape changes in pulsars
has dramatically changed our understanding of timing noise as well as the relationship between
the radio emission and the properties of the magnetosphere as a whole. Using Gaussian processes
we are able to model timing and emission variability using a regression technique that imposes
no functional form on the data. We revisit the pulsars first studied by Lyne et al. (2010). We not only confirm the emission and
rotational transitions revealed therein, but reveal further transitions and periodicities in 8 years of
extended monitoring. We also show that in many of these objects the pulse profile transitions between two
well-defined shapes, coincident with changes to the period derivative. With a view to the SKA and
other telescopes capable of higher cadence we also study the detection limitations of period derivative
changes.
\keywords{pulsars: general, stars: neutron, methods: data analysis, methods: statistial.}

\end{abstract}

\firstsection 
\section{Introduction}

Timing noise is low frequency structure in the timing residuals of pulsars.  It is characterised by a smooth variation of the pulsar's spin parameters about a best-fit timing model.  The residuals often show asymmetry in their minima and maxima and long term timing has revealed quasi-periodicities in the structure over years to decades (\cite[Hobbs et al. 2010]{hlk10}). Studies of the intermittent pulsar PSR B1931+24 (\cite[Kramer et al. 2006]{klo+06}) led to a significant advancement in the understanding of timing noise. The pulsar's timing noise was reduced by a factor of 20 if two values of the spin-down rate ($\dot{\nu}$) were used in the model - one when the pulsar was \emph{on} and another, slower rate, when the pulsar was \emph{off}.  This correlation between emission and $\dot{\nu}$, implies that the nulling behaviour is part of a global magnetospheric change in which the outward flow of current from the pulsar changes, resulting in emission variations and changes to the pulsar's braking torque.

Studies of this behaviour were extended to 17 further pulsars whose residuals were dominated by timing noise in \cite[Lyne et al. (2010)]{lhk+10} (hereafter, LHK).  Six of these sources exhibited systematic changes to the shape of the pulse profile (i.e., \emph{mode-switching}).  In all six cases, the value of $\dot{\nu}$ correlated with a particular pulse shape, implying that pulse nulling may be an extreme form of mode-switching.

Since LHK an additional $\sim$8 years of timing data have been acquired by Jodrell Bank on these 17 pulsars allowing new $\dot{\nu}$ and/or pulse shape transitions to be measured. Additionally, an updated observing backend that yields up to a factor of three improvement in the available time resolution, means we are potentially sensitive to previously undetectable variability. We have revisted the pulsars in LHK in order to search for new and previously undetected transitions in pulse shape and $\dot{\nu}$. In addition to this, we have undertaken simulations of variable $\dot{\nu}$ pulsars with a view to understanding the smallest $\dot{\nu}$ changes that accompany transitions, that are detectable with current timing programs.  

\section{Variable pulsar spin-down}

We have measured the spin-down evolution and profile stability of 17 pulsars using a Gaussian Processes Regression method technique (see \cite[Brook et al. 2014]{bkb+14}; \cite[Brook et al. 2015]{bjk+15}).  We have confirmed the $\dot{\nu}$ variability in all of these sources. We show three examples of $\dot{\nu}$ variations from our sample, in Figure 1. The $\dot{\nu}$ variations themselves are highly variable from pulsar to pulsar.  PSR B1540-06 appears to transition between two well defined values in a highly periodic manner. There is also a gradual secular change in the mean value of $\dot{\nu}$.  More complex structure is seen in PSR B1907+00 in which the $\dot{\nu}$ variations comprise clear major and minor peaks with a cycle time between major peaks of several years. Though this could be interpreted as the pulsar having three or more distinct $\dot{\nu}$ states, this behaviour is reminiscent of PSR B0919+06 in which the amount of time the pulsar spends in one of two $\dot{\nu}$ states changes over time resulting in consecutive peaks showing different amplitudes (see \cite[Perera et al. 2015]{psw+15}).

\begin{center}
\begin{figure*}[!htb]
    \includegraphics[width=0.3\columnwidth]{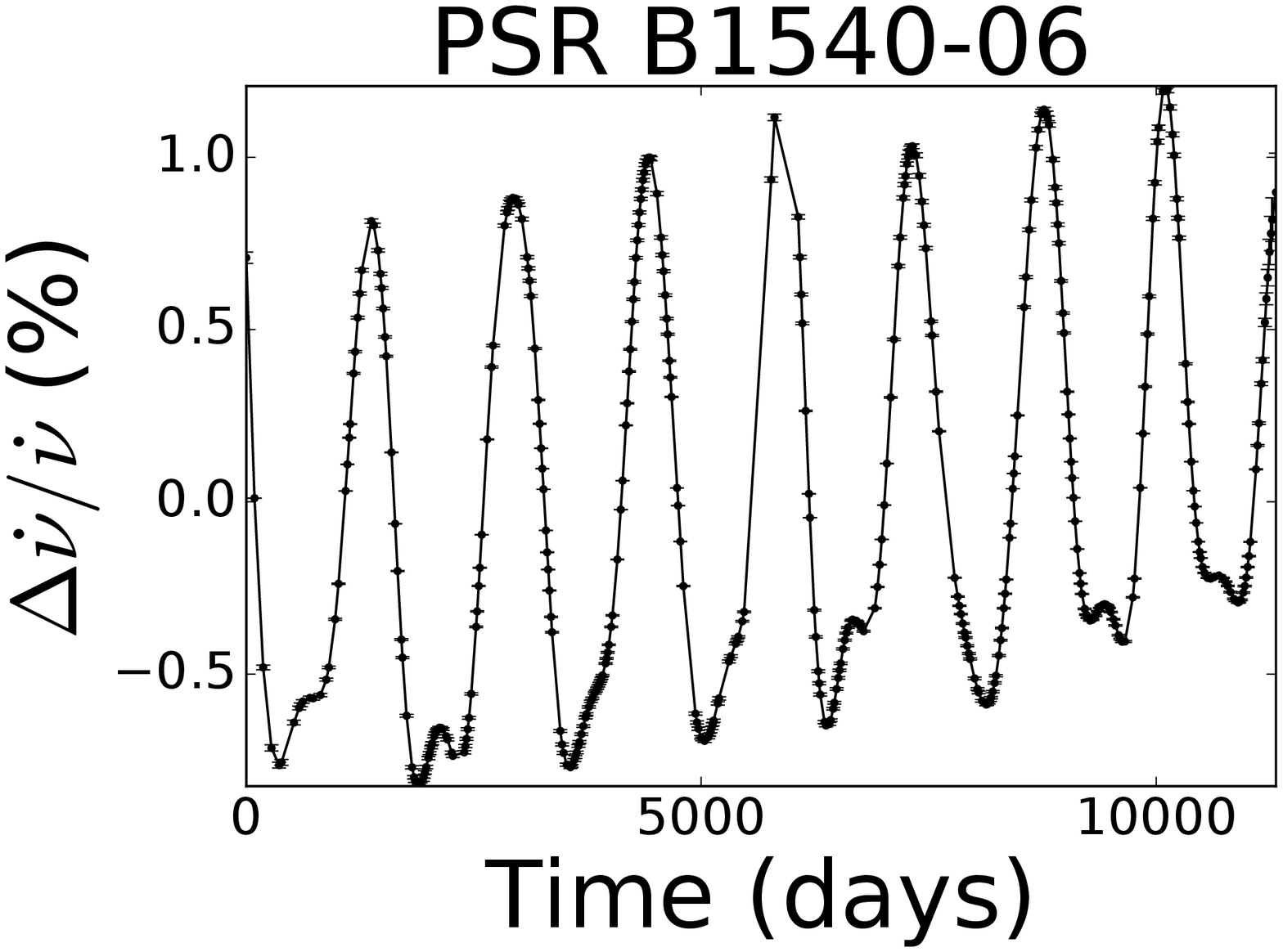}
    \includegraphics[width=0.3\columnwidth]{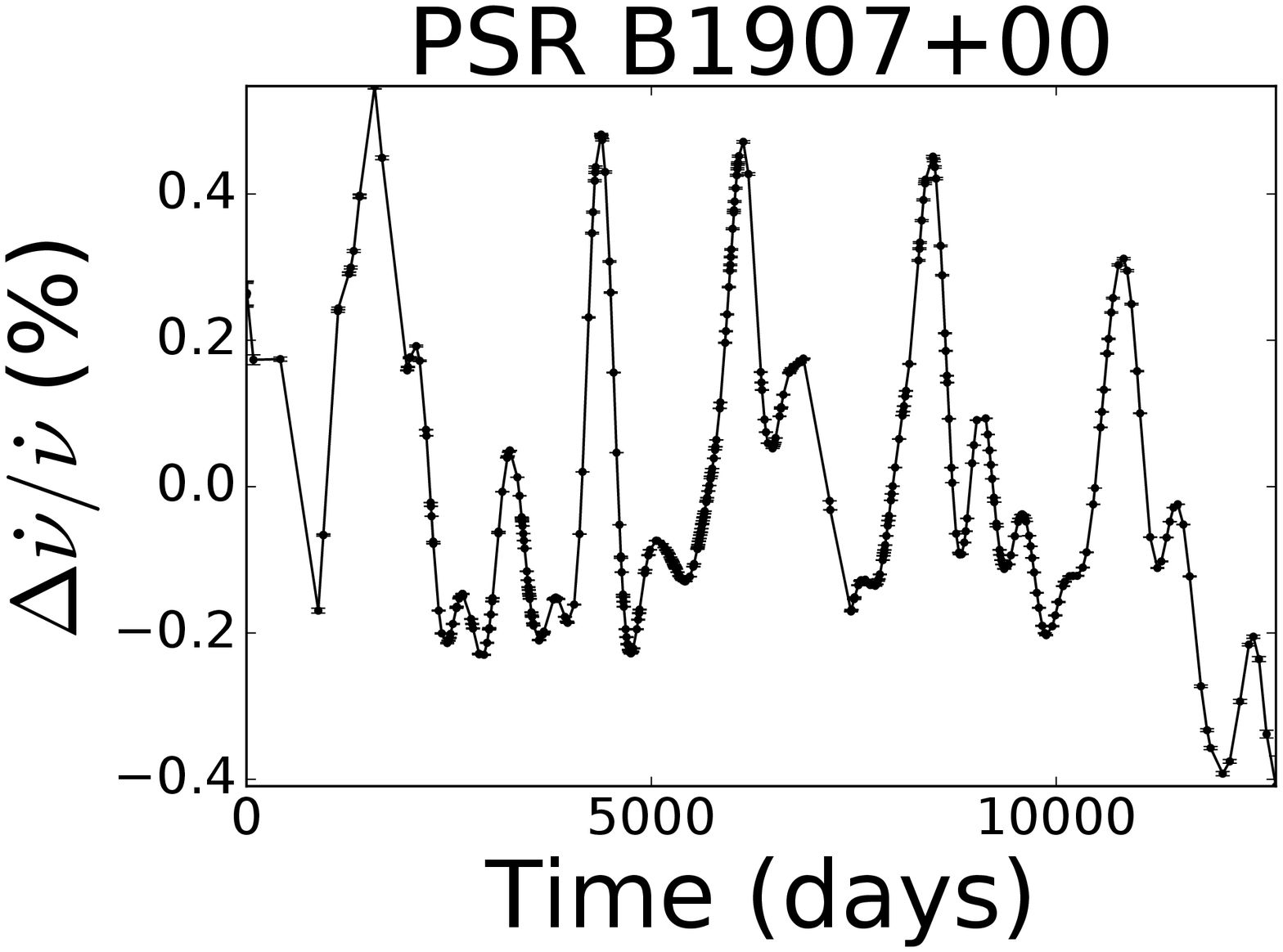} 
    \includegraphics[width=0.3\columnwidth]{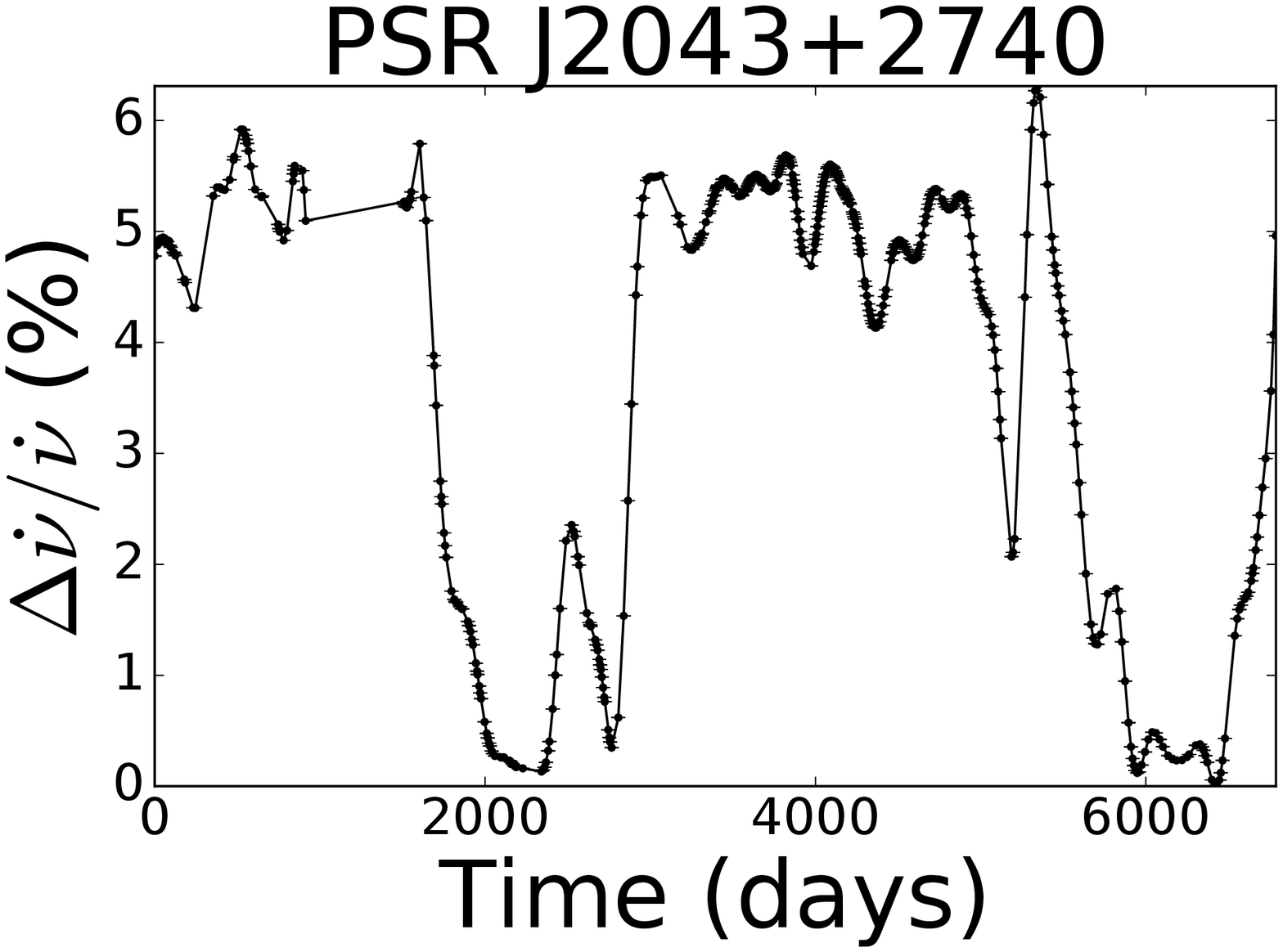}
    \label{nudots}
    \caption{The fractional change in $\dot{\nu}$ as a function of time for three pulsars from LHK, revisited in Shaw et al. (in prep). A downward deflection represents an increase in $\dot{\nu}$.}
\end{figure*}
\end{center}


Since LHK, we have detected and measured two further transitions in PSR J2043+2740. These are in addition to the two similar transitions observed in LHK.  The time between the two consective downward $\dot{\nu}$ transitions is approximately 10 years. We note that prior to two of these large transitions we observe apparently smaller amplitude transitions (i.e., prior to the first upward and second downward transitions) suggesting the possibility that greater mixing of the two states begins to occur as major transitions approach.

\section{$\dot{\nu}$-specific emission states}

LHK showed that six pulsars from their sample undergo mode-switching events contemporaneously with $\dot{\nu}$ transitions. We have confirmed all such emission-rotation correalation in these pulsars and revealed correlations in a number of other sources in the sample.  A particularly conspicuous example is exhibited by PSR J2043+2740 (Figure 2). This pulsar has been monitored at L-BAND by Jodrell Bank Observatory since 1996. At this time the pulse profile was characterised by a gradual rise followed by a slower, more gradual decline. In 2003 the pulsar underwent a mode-switch in which the trailing edge of the profile increased in amplitude as denoted by the bright regions in Figure 2. This profile change was accompanied by an increase in $\dot{\nu}$. The pulsar remained in this secondary mode until 2006 when both the profile and $\dot{\nu}$ returned to the primary mode.  We have observed a repetition of this transitioning behaviour starting in 2014 as a result of the continued monitoring of this source.

\vspace{-3mm}
\begin{figure}[!htb]
    \hspace{12mm} \includegraphics[width=0.9\columnwidth]{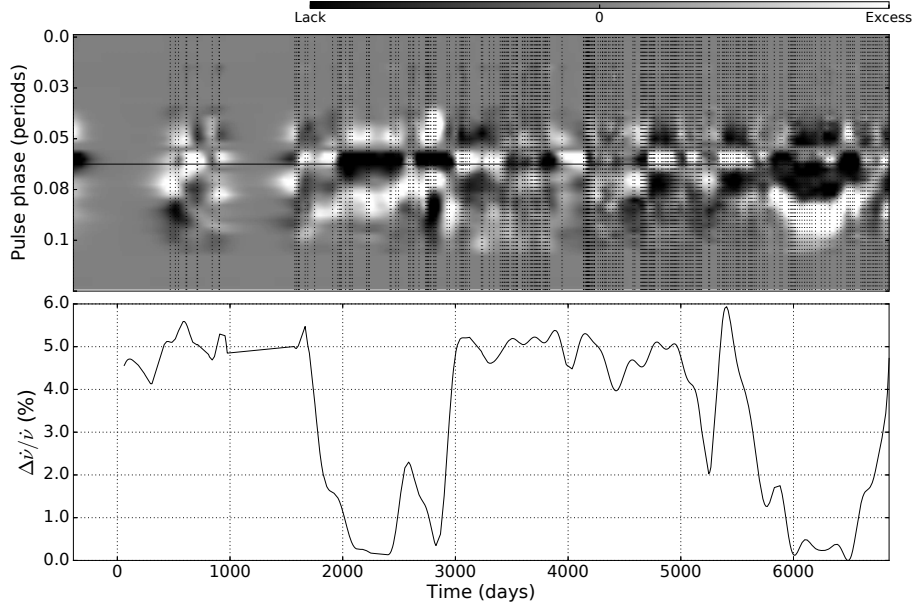}
   
    \label{2043}
    \caption{Comparison of the profile shape and the spin-down rate in PSR J2043+2740. In the upper plot, the pulse longitude increases from top to bottom. The horizontal line denotes the pulse peak. Light/dark regions denote an excess/lack of flux respectively with respect to the profile residuals, calculated by subtracting a high S/N template formed from all observations, from each individual profile. Vertical lines denote observation epochs.}
\end{figure}

This transitioning behaviour is consistent with a model in which a pulsar's spin-down is regulated by conditions in the magnetosphere. When the pulsar transitions to a mode in which current flow is enhanced, changes to the radio emission are observed as well as augmented braking due to additional torques on the pulsar.  There is no established consensus however, on why pulsar magnetospheres assume different configurations. Our results from the 17 pulsars in LHK will be described in \cite[Shaw et al. (2018b)]{ssb+18}.  

\section{Limits on $\Delta \dot{\nu}$-detection}

We have investigated the sensitivity of current long-term pulsar timing programs to variable $\dot{\nu}$ behavior. Large transitions in $\dot{\nu}$ that are separated by long intervals, such as those seen in PSR J2043+2740, have a strong effect on the timing residuals of pulsars by producing timing noise.  Depending on the available pulse time-of-arrival (TOA) precision, transitions that are small and more closely spaced in time may not cause the TOAs to deviate from the predictions of a simple timing model that contains a constant $\dot{\nu}$. In these cases, including transition parameters in a pulsar's timing model may not yield an improvement to the $\chi^2$ of the timing residuals and therefore the transitions are not resolvable.   We have taken a two-part approach to determine whether or not individual $\dot{\nu}$ transitions can be resolved for a wide range of transitioning scenarios.  Firstly, analytical limits have been derived which predict the extent to which a pulsar's timing residuals will depart from a simple slow-down model, due to unmodelled $\dot{\nu}$ transitions, taking into account TOA precision and observing cadence.  Secondly, pulse TOAs are simulated, into which a number of $\pm \Delta \dot{\nu}$ transitions are injected. We vary the time $\tau$ that elapses between these transitions and attempt to resolve them using standard pulsar timing techniques. We do this for a wide range of transition amplitudes and timescales, investigating the effects of cadence and TOA precision on detectability. Our simulations have shown that:


\begin{itemize}
  \item{A 1$\sigma$ probability of resolving transitions is consistent with our analytical limits.}
  \item{Where $\dot{\nu}$ and the profile shape jointly change, prior knowledge of the transition epochs, from monitoring the pulse profile, can improve the probability of resolving transitions in some cases.}
  \item{Currently attainable TOA precision may conceal a variety of $\dot{\nu}$ transitioning phenomena. More sensitive facilities such as the SKA may reveal timing noise in a greater number of sources, leading to an improved understanding of pulsar magnetospheres.}
  \item{Many pulsars, whose profiles change on very short timescales, may exhibit coincident $\dot{\nu}$ transitions but cadence is seldom sufficient for discrete $\dot{\nu}$ states to be resolved. New facilities such as CHIME and UTMOST, capable of improved observing cadence, will reveal new examples of sources that transition on timescales of days.}
  \item{When more sensitive telescopes are online, cadence should not be de-prioritised as frequent monitoring is required to resolve transitions in more rapidly switching sources.}
  \item{Some $\dot{\nu}$-transitioning scenarios may mimic glitches and vice-versa} 
\end{itemize}

Details of our simulations and findings are described in \cite[Shaw et al. (2018a)]{ssw18}. 

\vspace{-2mm}

\section{Conclusions}


Many pulsars appear to exhibit a variable $\dot{\nu}$. The nature of the variations is such that, at a given time, a pulsar assumes one of (usually) two discrete values of $\dot{\nu}$ and transitions between them on some characteristic timescale. The fact that in some cases, simultaneous changes to the pulse profile occurs, suggests each of the states corresponds to a particular magnetospheric configuration, though an explanation of why pulsar magnetospheres behave this way remains elusive. Though in only a handful of cases has a link been established between $\dot{\nu}$ and emission, timing noise is widespread across the pulsar population, suggesting that transitioning magnetospheres are ubiquitous.  

We have revisited the pulsars in LHK with an improved time resolution and baseline. We have confirmed the $\dot{\nu}$ and profile variations therein and revealed new examples of this behaviour. We have also undertaken simulations of $\dot{\nu}$-transitioning pulsars and find that with current observing setups we remain potentially insensitive to a rich diversity of magnetospheric phenomena. New generations of telescopes have the potential to reveal much of this behaviour, if sensitivity and cadence are appropriately configured, leading to an improved understanding of pulsar magnetospheres.

\end{document}